# DAFNE2: LATTICE FOR A 2 GeV ELECTRON COLLIDER

G. Benedetti
LNF-INFN, Frascati, Italy

*Abstract*

Within the general frame of the upgrade of the collider DAΦNE to the neutron-antineutron energy threshold (2–2.5 *GeV* c.m.), modifications and different options for the accelerator design are discussed and a possible lattice for the Main Rings at 1 *GeV* is presented.

The luminosity required by the experiments for such a *light quark factory* is of the order of few $10^{31}$ cm$^{-2}$ s$^{-1}$, easily achievable in the particle factory era.

## ENERGY UPGRADE OF THE FRASCATI FACTORY

The experimental program of DAFNE2 (Double Annular Frascati e$^+$e$^-$ factory for Nice Experiments at 2 *GeV*) can be carried out with the FINUDA detector with a solenoid field of 0.3 *T*.

Many hardware components from DAΦNE can be used for DAFNE2 without modifications: the vacuum chamber, all the electromagnetic quadrupoles and sextupoles, the RF system, the feedback, the diagnostic and the vacuum systems. However eight new stronger bending dipoles per ring and four superconducting quadrupoles in the experimental Interaction Region (IR2) are needed for the energy upgrade.

Concerning the injection, two options are possible:
- injection off-energy at 0.51 *GeV* with the current system (linac, damping ring and transfer lines) performing the energy ramping in the Main Rings[1];

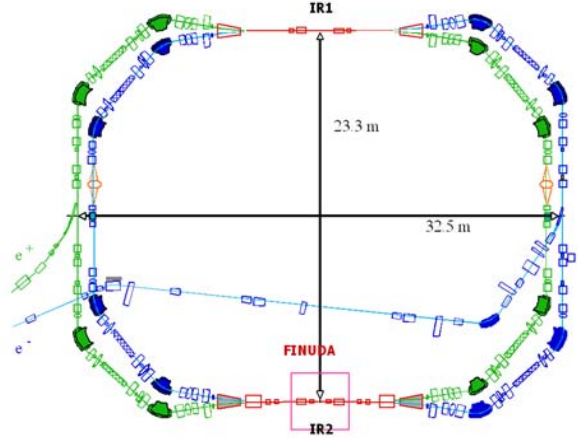

Figure 1: Main Rings layout.

- injection directly on-energy (1–1.1 *GeV*) adding new accelerating structures to the linac [2]. In this case the damping ring is not used and some modification to the transfer lines is needed.

*Machine Parameters*

The experimental luminosity requirements are not critical for DAFNE2. This allows choosing the machine parameters with enough freedom, exploiting our commissioning know-how. We have worked out new parameters at beam collision energy of 1 *GeV*.

Choosing an emittance $\varepsilon = 0.5 \cdot 10^{-6}$ *rad m* compatible with the ring aperture and a vertical beta function at the Interaction Point (IP) and a coupling factor already achieved (Table 1), the linear tune shift comes out $\xi_x / \xi_y$ = 0.014 / 0.024, below the limit achieved by DAΦNE.

The horizontal crossing angle at the interaction point is ±15 *mrad*, corresponding to a Piwinsky angle:

$$\phi = \theta_x^* \frac{\sigma_L}{\sigma_x^*} = 0.22$$

which has already been exceeded in the existing factories.

The chosen number of bunches is 30 so that, leaving the harmonic number $h = 120$ unchanged, we can inject both electrons and positrons out of collision at 0.51 *GeV* and collide the beams by performing a fast RF phase jump [3] as the beams have been accelerated to the collision energy.

In the on-energy injection option, injection in collision is also allowed and in this case different multibunch configurations can be used.

Once fixed such parameters (Table 1), a luminosity of $1 \cdot 10^{32}$ s$^{-1}$ cm$^{-2}$ is straightforward to achieve with 15 *mA* per bunch corresponding to a beam current of 0.45 *A*.

| | | |
|---|---|---|
| Energy $E_0$ | 1.0 | *GeV* |
| Luminosity L | $1 \cdot 10^{32}$ | s$^{-1}$ cm$^{-2}$ |
| Circumference C | 97.69 | *m* |
| Emittance ε | $0.5 \cdot 10^{-6}$ | *rad m* |
| Coupling $\kappa = \varepsilon_x / \varepsilon_y$ | 0.003 | |
| Beta functions at IP $\beta_x^* / \beta_y^*$ | 1.5 / 0.025 | *m* |
| Crossing angle at IP $\theta_x^*$ | ±15 | *mrad* |
| Bunch width at IP $\sigma_x^* / \sigma_y^*$ | 0.95 / 0.008 | *mm* |
| Bunch natural length $\sigma_z$ | 13.9 | *mm* |
| Linear tune shift $\xi_x / \xi_y$ | 0.014 / 0.024 | |
| Betatron tunes $\nu_x / \nu_y$ | 5.15 / 5.21 | |
| Momentum compaction $\alpha_c$ | 0.012 | |
| Number of bunches | 30 | |
| Beam current $I_{tot}$ | 450 | *mA* |

Table 1: DAFNE2 Parameters

# MAIN RING DESIGN

The design of the two Main Rings is unchanged (Figure 1): two different rings for positrons and electrons with two 10 *m* long Interaction Regions where the opposite beams travel in the same vacuum chamber. Four wigglers per ring are installed in the arcs.

The experimental detector FINUDA is housed in the Interaction Region 2 (IR2) and two quadrupole doublets are installed in the opposite Interaction Region (IR1): a lattice already used before 2003 when only the KLOE detector was installed [4].

## New Dipoles

Currently eight dipoles are installed in each ring: four 0.99 *m* long magnets with a 40.5° bending angle (two are sector type and two rectangular) and four 1.21 *m* long with a 49.5° angle (two sector and two rectangular). The bending field in the current dipoles at 0.51 *GeV* is 1.2 *T* and the maximum field they can reach is 1.7 *T*, insufficient to reach 1 *GeV*: new stronger dipoles are needed.

The existing vacuum chamber imposes constraints on the dipole geometry: the new magnets can be 10% longer and all sector type to have the maximum allowed magnetic length. With a different shape of the polar shoe made of *permendur* and with the gap height reduced from 4 *cm* to 3.7 *cm*, we can achieve the needed field of 2.3 *T*. More work and simulations are in progress to study the features and the quality of such magnets [5].

## Wiggler OFF/ON options: Synchrotron Radiation

Synchrotron radiation per turn depends on the energy and on the bending radius in dipoles and wigglers as:

$$U_0 = C_\gamma \frac{E^4}{2\pi} \oint \frac{ds}{\rho^2}$$

Four wiggler magnets per ring are currently used in DAΦNE to double the lower the damping times and to modulate the natural emittance of the beam.

In the wiggler ON option the wiggler field is constant from 0.51 to 1 *GeV*, because in the 0.51 *GeV* lattice it is in saturation and at 1 *GeV* the radiation from wigglers is enough to yield short damping times (Table 2). The drawback of this option is the difficulty to control the nonlinearities generated by the wigglers during the energy ramping.

On the other hand the wiggler OFF option is not much performing at low energy, but it is preferable in the on-energy injection case because at 1 *GeV* damping times are enough short also without wigglers.

| **Wigglers OFF/ON** | | 0.51 GeV | 1 GeV |
|---|---|---|---|
| $U_0$ | (*keV/turn*) | 4.3 / 9.3 | 64.0 / 83.5 |
| $\tau_x$ | (*ms*) | 68 / 40 | 11 / 8.6 |
| $\tau_E$ | (*ms*) | 41 / 31 | 5.0 / 3.5 |

Table 2: Energy loss with and without wigglers

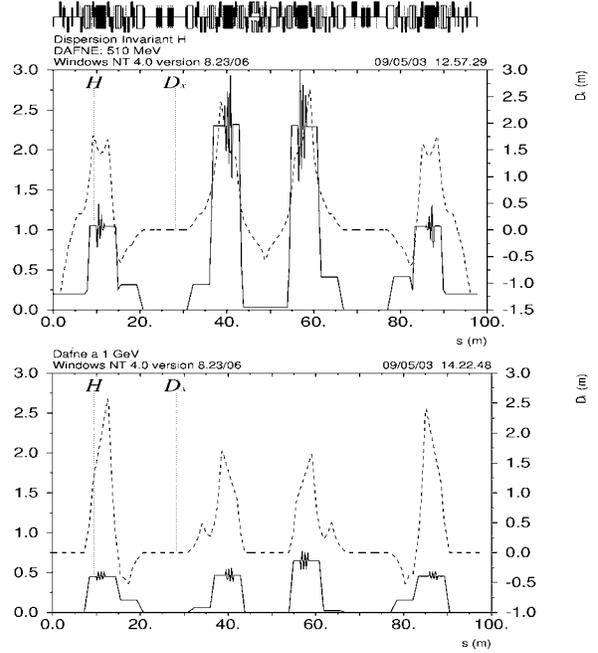

Figures 2 and 3: the dispersion function and the invariant H at low (above) and high energy (below).

The emittance in electron storage rings depends on the second power of the energy according to:

$$\varepsilon = C_q \gamma^2 \frac{<\mathcal{H}/|\rho|^3>}{J_x <1/\rho^2>}$$

In the off-energy injection option to have emittance values compatible with lifetime and aperture requirements at different energies the dispersion invariant *H*:

$$\mathcal{H} = \gamma_x D^2 + 2\alpha_x DD' + \beta_x D'^2$$

must be fairly reduced by changing the Twiss parameters in the arcs during the energy ramping (Figures 2 and 3).

## Optics

Two very similar lattices wiggler OFF/ON at 1 *GeV* have been calculated (Figures 4 and 5) with the following main features:

- only one low beta insertion in IR2, where the experimental detector is housed and the positron and electron beams collide with a horizontal angle of ±15 *mrad*;
- in IR1, four FDDF quadrupoles and a 'detuned' vertical beta configuration (Figure 6) allow to separate the beam trajectories with a vertical bump of ±1 *cm*;
- horizontal and vertical beta functions in the four achromat arcs, where the dispersion is higher, are separated to correct both horizontal and vertical natural chromaticity with chromatic sextupoles;
- the horizontal beta function and the dispersion are shaped in such a way that the natural emittance does not change from 0.51 to 1 *GeV* as explained in the section above.

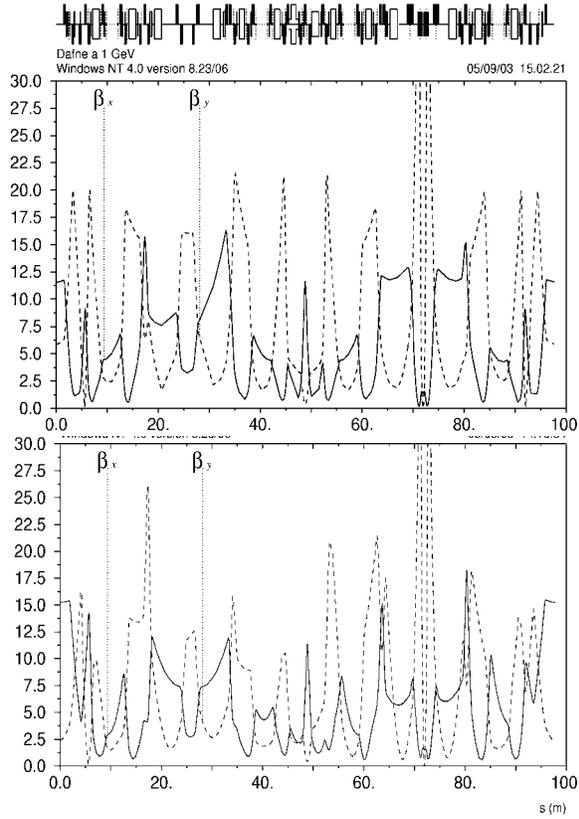

Figures 4 and 5: Main Ring wiggler ON (above) and wiggler OFF (below) beta functions calculated by MAD.

## Interaction Region 2

The low beta insertion is realized with four FDDF quadrupoles housed inside the experimental detector (Figure 7). Since they must be powered for variable beam energy, the only solution to fit them inside the detector is developing superconducting quadrupoles. Two further doublets outside the detector are realized with conventional quadrupoles.

The FINUDA solenoid field integral of 0.3 $T$ x 2.4 $m$ at the collision energy rotates the beam by an angle of 6° around the longitudinal axis. The coupling at the Interaction Point and outside IR2 can be corrected with the rotating frame method [6] as in DAΦNE: each quadrupole outside is rotated around its longitudinal axis following the rotation of the beam and two compensating solenoids 0.36 $T$ x 1 $m$ provide cancellation of coupling outside the IR. Another possibility, instead of rotating the quadrupoles inside the detector, is to use superconducting quadrupoles with independent normal and skew windings as developed at CESR [7]. The solenoid fields and the quadrupole rotation angles are fixed. At the injection energy of 0.51 $GeV$ the non vanishing coupling from mismatched quadrupole rotation angles can be useful to have a long beam lifetime during the ramping and if necessary it can be controlled with the skew quadrupoles installed along the ring.

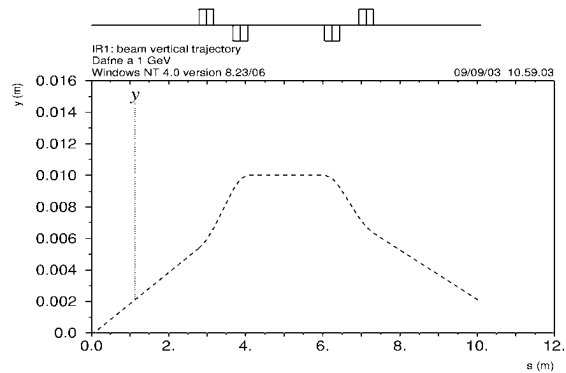

Figure 6: vertical beam trajectory $y$ along IR1.

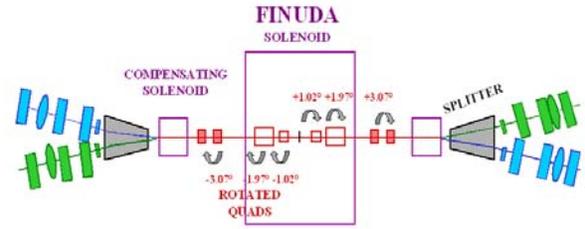

Figure 7: Interaction Region 2 layout.

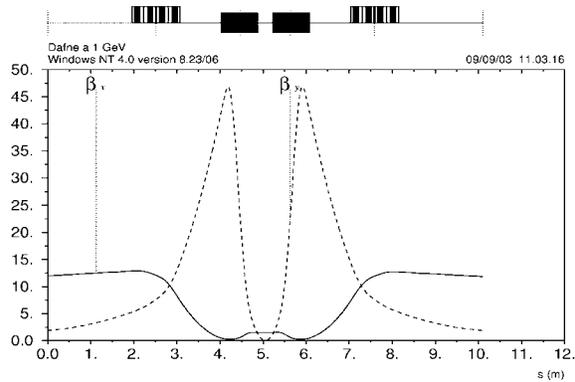

Figure 8: beta functions along IR2.

## RF Parameters and longitudinal bunch distribution

The DAΦNE RF cavity cooling system can withstand a maximum accelerating field of 350 $kV$, corresponding to a RF power loss of 35 $kW$ on the cavity walls, while the maximum RF power the klystron can supply is 150 $kW$. The RF power to be delivered to the beam is given by $P_{beam}= V_{loss} I_{beam}\approx$ 40.5 $kW$, assuming 90 $keV/turn$ of total losses (including the parasitic ones). Since the required accelerating voltage is 250 $kV$ corresponding to a RF wall dissipation of ≈17.5 $kW$, the existing RF system is completely compatible with the requirements.

Bunch lengthening has been estimated by performing a multiparticle tracking. Using the impedance estimates and corresponding wake fields calculated for the present vacuum chamber [8] in the turbulent microwave threshold calculations and in the bunch lengthening simulations, the rms bunch length increases from the natural value of 13.9 *mm* (Figure 9) to only 15.9 *mm* at 15 *mA* per bunch, while the energy spread remains constant indicating that the microwave instability threshold is not reached at the nominal bunch current.

Table 3: RF System and Bunch Parameters

| | |
|---|---|
| RF peak voltage $V_{RF}$ | 250 *kV* |
| RF frequency $f_{RF}$ | 368.26 *MHz* |
| Energy loss $U_{rad}+U_{paras}$ | 83.5 +6.5 *KeV/turn* |
| RF power $P_{beam}+P_{wall}$ | 40.5 + 17.5 *kW* |
| Synchr. frequency $f_{syn}$ | 11.7 *kHz* |

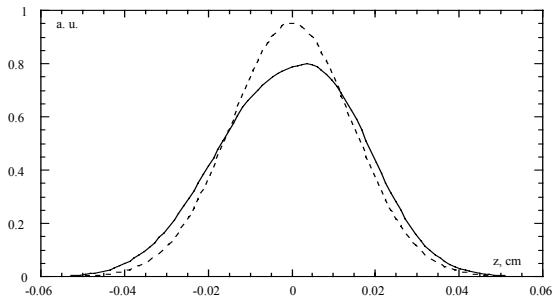

Figure 9: Charge density bunch distribution at zero current (dashed line) and at 15 *mA/bunch* (solid line).

## *Lifetime and background*

Background and beam lifetime at DAΦNE are strongly dominated by Touschek scattering. Touschek lifetime is a complicated function of machine parameters: at the larger energy and RF voltage of DAFNE2 it will be less critical than at the present energy. In fact with the parameters in Tables 1 and 3 $\tau_{tou}$ comes out to be 650 *min* as calculated by MAD8 with longitudinal acceptance dominated by RF. Further quantitative simulations will be done with the programs developed and used for DAΦNE that consider the physical aperture of the vacuum chamber along the rings.

## *Vacuum System*

Present layout can withstand the new configuration. In fact in DAFNE2 (0.45 *A* and 1 *GeV*) synchrotron radiated photon flux is $1.8·10^{20}$ *ph/s* corresponding to a power of 38 *kW*, while the existing vacuum chamber is designed for a synchrotron radiation power of 50 *kW*.

## *Feedback System*

No change is needed for the transverse feedback if the betatron tunes stay constant during the energy ramping. The longitudinal feedback can follow the synchrotron frequency variation in a large range with eight synchronizable filters. Timing is not critical within a synchronous RF phase up to 100 *ps* (the RF phase is 70 *ps* at 250 *kV*).